# Iterative-Promoting Variable Step-size Least Mean Square Algorithm For Adaptive Sparse Channel Estimation

Beiyi Liu, Guan Gui and Li Xu

Department of Electronics and Information Systems, Akita Prefectural University

Yurihonjo 015-0055, Japan

Emails: lby31411@gmail.com , {guiguan, xuli}@ akita-pu.ac.jp

*Abstract*—Least mean square (LMS) type adaptive algorithms have attracted much attention due to their low computational complexity. In the scenarios of sparse channel estimation, zero-attracting LMS (ZA-LMS), reweighted ZA-LMS (RZA-LMS) and reweighted $\ell_1$-norm LMS (RL1-LMS) have been proposed to exploit channel sparsity. However, these proposed algorithms may hard to make tradeoff between convergence speed and estimation performance with only one step-size. To solve this problem, we propose three sparse iterative-promoting variable step-size LMS (IP-VSS-LMS) algorithms with sparse constraints, i.e. ZA, RZA and RL1. These proposed algorithms are termed as ZA-IPVSS-LMS, RZA-IPVSS-LMS and RL1-IPVSS-LMS respectively. Simulation results are provided to confirm effectiveness of the proposed sparse channel estimation algorithms.

## I. INTRODUCTION

The demand for high-speed wireless communications has been increasing daily. Broadband signal transmission is one of indispensable techniques in next wireless communication systems [1]. In the broadband channels, most of the energy of the finite impulse response (FIR) is concentrated in a small fraction of its duration which is usually called sparse channel, due to the fact that broadband signal is transmitted over frequency-selective fading channel [2]. Hence, accurate channel estimation is required for coherent detection.

To estimate the sparse channel, many sparse adaptive channel estimation algorithms have been proposed [1][3]–[6]. By using the invariable step-size(ISS), [1] proposed zero-attracting least mean square (ZA-LMS), reweighted ZA-LMS (RZA-LMS) and reweighted $\ell_1$-norm LMS (RL1-LMS) to exploit the channel sparsity. However, these sparse LMS algorithms have the common shortcoming which cannot make tradeoff between convergence speed and estimation performance by ISS. On the one hand, utilizing a smaller step-size can achieve a better estimation performance but scarifying the convergence speed. On the other hand, utilizing a larger step-size can improve convergence speed but deteriorating the estimation performance.

Motivated by ISS-LMS-type algorithms, this paper proposes iterative-promoting variable step-size LMS (IPVSS-LMS) algorithms for estimating sparse channels. The proposed algorithms have a larger step-size in the initial stage, and then step-size is iteratively reduced. To achieve steady-state solution, the VSS is also bounded by a threshold. Hence, the proposed sparse IPVSS-LMS algorithms can balance the instantaneous estimation error and convergence speed. Since the threshold is adopted for VSS, the proposed sparse IPVSS-LMS algorithms do not scarify the additional convergence speed. Several representative simulations are conducted to confirm the proposed algorithms with respect to different scenarios.

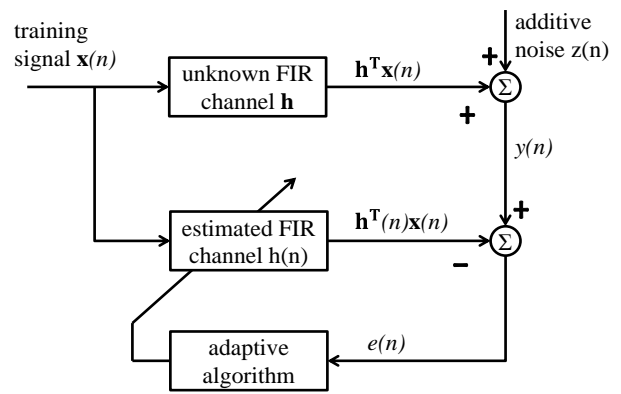

Fig. 1.   ASCE for broadband communication systems

The remainder of the rest paper is organized as follows. A system model is described and sparse ISS-LMS algorithms are reviewed in Section II. In Section III, ZA-IPVSS-LMS, RZA-IPVSS-LMS and RL1-IPVSS-LMS are proposed. The simulation results are presented in Section IV. Finally, this paper is concluded in Section V.

## II. PROBLME FORMULATION

Adaptive channel estimation method for estimating wireless channels is illustrated in Fig.1. Assume that the training signal $x(n)$ transmit over the sparse channel $\boldsymbol{h} = [h_0, h_1, ..., h_{N-1}]^T$ which is supported only by $K$ non-zero coefficients ($K \ll N$). Then the output signal $y(n)$ is obtained as

$$y(n) = \boldsymbol{h}^T \boldsymbol{x}(n) + z(n) \tag{1}$$

where $\boldsymbol{x}(n) = [x(n), x(n-1), ..., x(n-N+1)]^T$ denotes the N-length vector of training signal $x(n)$; $z(n)$ is random additive white Gaussian noise $X \sim N(0,1)$ which is independent of $x(n)$. The objective of the adaptive filter is to estimate the unknown sparse channel $\boldsymbol{h}(n)$ by utilizing training signal $x(n)$ and output signal $y(n)$. Then the $n$-th instantaneous estimation error $e(n)$ can be written as

$$e(n) = y(n) - \tilde{\boldsymbol{h}}(n)^T \boldsymbol{x}(n) \tag{2}$$

where $\tilde{\boldsymbol{h}}(n)$ is the estimated channel. According to Eq. (2), the cost function of the LMS algorithm can be written as

$$L(n) = \frac{1}{2}e^2(n) \tag{3}$$

By derivating Eq. (3) with respect to $\tilde{\boldsymbol{h}}(n)$, the update equation of LMS is obtained as

$$\tilde{\boldsymbol{h}}(n+1) = \tilde{\boldsymbol{h}}(n) + \mu \frac{\partial L(n)}{\partial \tilde{\boldsymbol{h}}(n)} \tag{4}$$
$$= \tilde{\boldsymbol{h}}(n) + \mu e(n) \boldsymbol{x}(n)$$

where $\mu$ is a step size of gradient descend step-size.

One can be observed that, ISS-LMS in Eq. (4) does not exploit the channel sparsity. To exploit the channel sparisity, sparse ISS-LMS is proposed as

$$\tilde{\boldsymbol{h}}(n+1) = \underbrace{\underbrace{\tilde{\boldsymbol{h}}(n) + \mu e(n) \boldsymbol{x}(n)}_{LMS} - \lambda Sparse\ Penalty}_{Sparse\ LMS} \tag{5}$$

where $\lambda$ denotes regularization parameter, which can balance the estimation error term and channel sparsity exploitation. We next review three sparse ISS-LMS algorithms. They are ZA-LMS, RZA-LMS and RL1-LMS.

A. ZA-LMS. The cost function of ZA-LMS is given by:

$$L_{ZA}(n) = \frac{1}{2}e^2(n) - \lambda_{ZA} \left\| \tilde{\boldsymbol{h}}(n) \right\|_1 \tag{6}$$

where $\lambda_{ZA}$ is the weight associated with the penalty term and $\|\cdot\|_1$ denotes the $\ell_1$-norm. Then the updated equation of ZA-LMS is derived as:

$$\tilde{\boldsymbol{h}}(n+1) = \tilde{\boldsymbol{h}}(n) + \mu \frac{\partial L_{ZA}(n)}{\partial \tilde{\boldsymbol{h}}(n)} \tag{7}$$
$$= \tilde{\boldsymbol{h}}(n) + \mu e(n) \boldsymbol{x}(n) - \rho_{ZA} \operatorname{sgn}\left(\tilde{\boldsymbol{h}}(n)\right)$$

where $\rho_{ZA} = \mu \lambda_{ZA}$ and $\operatorname{sgn}(\cdot)$ expressed as follows:

$$\operatorname{sgn}(h) = \begin{cases} 1, & h > 0 \\ 0, & h = 0 \\ -1, & h < 0 \end{cases}$$

Sparse penalty term of ZA-LMS can attract every channel coefficient to be zero, hence, convergence speed can be improved by ZA-LMS.

B. RZA-LMS. The cost function of RZA-LMS is given by:

$$L_{RZA}(n) = \frac{1}{2}e^2(n) - \lambda_{RZA} \sum_{i=1}^{N} \log\left(1 + \varepsilon |h_i|\right) \tag{8}$$

where $\lambda_{RZA} > 0$ is the weight associated with the penalty term and $\varepsilon > 0$ is the reweight as a positive threshold. Then i-th channel coefficient $\tilde{h}_i(n)$ is given as:

$$\tilde{h}_i(n+1) = \tilde{h}_i(n) + \mu \frac{\partial L_{RZA}(n)}{\partial h_i(n)} \tag{9}$$
$$= \tilde{h}_i(n) + \mu e(n) x_i(n) - \rho_{RZA} \frac{\operatorname{sgn}\left(\tilde{h}_i(n)\right)}{1 + \varepsilon \left|\tilde{h}_i(n)\right|}$$

where $\rho_{RZA} = \mu \varepsilon \lambda_{RZA}$. Equation 9 can be expressed easily in the vector form as:

$$\tilde{\boldsymbol{h}}(n+1) = \tilde{\boldsymbol{h}}(n) + \mu e(n) \boldsymbol{x}(n) - \rho_{RZA} \frac{\operatorname{sgn}\left(\tilde{\boldsymbol{h}}(n)\right)}{1 + \varepsilon \left|\tilde{\boldsymbol{h}}(n)\right|} \tag{10}$$

Sparse penalty term of RZA-LMS can attract channel coefficient whose magnitudes are comparable to $1/\varepsilon_{RZA}$ to be 0.

C. RL1-LMS. The cost function is given by:

$$L_{RL1}(n) = \frac{1}{2}e^2(n) - \lambda_{RL1} \left\| \boldsymbol{f}(n) \tilde{\boldsymbol{h}}(n) \right\|_1 \tag{11}$$

where $\lambda_{RL1}$ is the weight associated with the penalty term and vector $f(n)$ is set as:

$$f(n) = \begin{bmatrix} f_0(n) \\ f_1(n) \\ \vdots \\ f_{N-1}(n) \end{bmatrix} = \begin{bmatrix} \frac{1}{\delta + |h_0(n-1)|} \\ \frac{1}{\delta + |h_1(n-1)|} \\ \vdots \\ \frac{1}{\delta + |h_{N-1}(n-1)|} \end{bmatrix} \quad (12)$$

where $\delta$ is some positive number and hence $f_i(n) > 0$ for $i = 0,1,...,N-1$. Then the updated equation of RL1-LMS algorithm is derived as:

$$\begin{aligned}\tilde{h}(n+1) &= \tilde{h}(n) + \mu \frac{\partial L_{RL1}(n)}{\partial h(n)} \\ &= \tilde{h}(n) + \mu e(n)x(n) \\ &\quad - \mu \lambda_{RL1} \operatorname{sgn}(f(n)\tilde{h}(n))\tilde{h}(n) \\ &= \tilde{h}(n) + \mu e(n)x(n) - \rho_{RL1} \frac{\operatorname{sgn}(\tilde{h}(n))}{\delta + |\tilde{h}(n-1)|}\end{aligned} \quad (13)$$

It can be easily proved that RL1-LMS has a stronger attraction than RZA-LMS because of the different penalty term [4].

## III. PROPOSE IPVSS-LMS ALGORITHMS

It is well known that the step-size is a critical parameter which determines the convergence speed, estimation performance and computational complexity [7], and decreasing the step-size with increasing of the iterative is an effective method. The updated equation of IPVSS-LMS for estimating channel can be pressed as:

$$\begin{aligned}\tilde{h}(n+1) &= \tilde{h}(n) + \mu(n) \frac{\partial L(n)}{h(n)} \\ &= \tilde{h}(n) + \mu(n)\{e(n)x(n) - Sparse\ Penalty\}\end{aligned} \quad (14)$$

where $\mu(n)$ is devised as

$$\mu(n) = \begin{cases} \mu/n, & if\ \mu(n) \geq \varphi \\ \varphi, & if\ \mu(n) < \varphi \end{cases} \quad (15)$$

where $n$ is the number of the iterative and $\varphi$ is a hard threshold to ensure the convergence when $\mu$ is enough small. Then we substitute Eq. (15) into Eq. (7,10,13) respectively, and achieve the updated equation as:

A. *ZA-IPVSS-LMS:*

$$\begin{aligned}\tilde{h}(n+1) &= \tilde{h}(n) \\ &\quad + \mu(n)\{e(n)x(n) - \lambda_{ZA}\operatorname{sgn}(\tilde{h}(n))\}\end{aligned} \quad (16)$$

B. *RZA-IPVSS-LMS:*

$$\begin{aligned}\tilde{h}(n+1) &= \tilde{h}(n) \\ &\quad + \mu(n)\left\{e(n)x(n) - \varepsilon\lambda_{RZA} \frac{\operatorname{sgn}(\tilde{h}(n))}{1 + \varepsilon_{RZA}|\tilde{h}(n)|}\right\}\end{aligned} \quad (17)$$

C. *RL1-IPVSS-LMS:*

$$\begin{aligned}\tilde{h}(n+1) &= \tilde{h}(n) \\ &\quad + \mu(n)\left\{e(n)x(n) - \lambda_{RL1} \frac{\operatorname{sgn}(\tilde{h}(n))}{\delta + |\tilde{h}(n-1)|}\right\}\end{aligned} \quad (18)$$

where the magnitudes of $\mu(n)$ is just decided by $1/n$ and $\varphi$, hence the calculation is very simple. There are two steps as shown in Fig.2:

*Step I:* $\mu(n) = \mu/n$ when $\mu/n \geq \varphi$: A fast convergence speed is achieved.

*Step II:* $\mu(n) = \varphi$ when $\mu/n < \varphi$: The estimation performance can be directly controlled by the hard threshold $\varphi$ which we set.

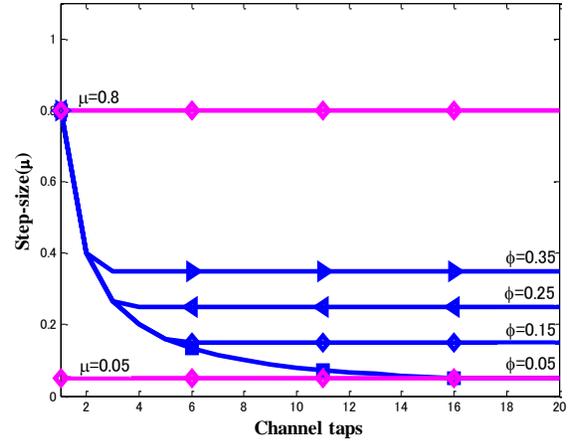

Fig. 2. Step-size of VSS-LMS algorithm against threshold (pink line denotes the ISS.

## IV. COMPUTER SIMULATION

In this section, the estimation performance of IPVSS-LMS for estimating channel is verified and compared with two ISS-LMS algorithms for estimating channel. One is used the step-size 0.005 and another is used the step-size 0.0005 which equals the hard threshold $\varphi$. The length of channel vector $h$ is set as 128 with $K(K \in \{4,8,12\})$ non-zero coefficients. The values of dominant channel taps follow random Gaussian distribution and the positions of non-zero coefficients are random. The training signal is Pseudo-random binary sequence whose coefficient is only 1 and -1. The received signal-to-noise (SNR) is defined as $10\log(E_s/\sigma_n^2)$, where $E_s = 1$ is the unit transmission power. In this paper, we set SNR as 5, 10 and 15 to compare each other. The noise is set as additive white Gaussian. All of the simulation parameters are listed in Table. I.

The estimation performance is evaluated by average mean square error (MSE) which is defined by:

$$MSE\{\tilde{h}(n)\} = 10\log_{10} E\left[\|h - \tilde{h}(n)\|_2^2\right] \quad (19)$$

| Parameters | | ISS | VSS |
|---|---|---|---|
| Training signal $x(n)$ | | \multicolumn{2}{c|}{Pseudo-random binary sequence} |
| Noise $z(n)$ | | \multicolumn{2}{c|}{Additive white Gaussian $z(n) \sim N(0,1)$} |
| Channel length | | \multicolumn{2}{c|}{$N = 128$} |
| No. of non-zero coefficient | | \multicolumn{2}{c|}{$K \in \{4,8,12\}$} |
| SNR | | \multicolumn{2}{c|}{$SNR \in \{5,10,15\}$ dB} |
| $\mu_{upper}$ | | 0.005 | $\begin{cases}1/n, if\ \mu(n) \geq \varphi\\ \varphi, if\ \mu(n) < \varphi\end{cases}$ |
| $\mu_{lower}$ | | $\varphi = 0.0005$ | |
| Regularization parameters | ZA-LMS | \multicolumn{2}{c|}{$\lambda_{ZA} = 0.004$} |
| | RZA-LMS | \multicolumn{2}{c|}{$\lambda_{RZA} = 0.002$} |
| | | \multicolumn{2}{c|}{$\varepsilon = 20$} |
| | RL1-LMS | \multicolumn{2}{c|}{$\lambda_{RL1} = 0.004$} |
| | | \multicolumn{2}{c|}{$\delta = 0.05$} |

TAB.1. Simulation parameters

We compare the performance of proposed channel estimators using 1000 independent Monte-Carlo runs for averaging.

As shown in Figs.3-8, in the case of different SNR regimes, VSS-LMS for estimating sparse channel which is proposed always has better estimation performance than ISS-LMS for estimating sparse channel with step-size $\mu_{upper}$, achieves higher convergence speed than ISS-LMS for estimating channel with step-size $\mu_{lower}$, and almost doesn't sacrifice the computational complexity.

Let us take the Fig. 6 for example to further illustrate the advantages of the proposed algorithms. In the case of SNR is 5dB, the number of non-zero coefficient is 6, they are compared with two groups of the performance curves of ISS-LMS for estimating sparse channel with different step-sizes (0.005 and 0.0005). Step I, the proposed algorithms have a high speed convergence speed as same as ISS-LMS with step-size 0.005. When ISS-LMS with step-size 0.005 reach steady-state, the proposed algorithms continue to decline until $\mu(n) = 0.0005$. Step II, the steady-state performance curves of the proposed algorithms are same as ISS-LMS with step-size 0.0005. While it is obviously that the proposed algorithms have a higher convergence speed than ISS-LMS with step-size 0.0005. In other words, the proposed algorithms have the convergence speed of ISS-LMS with step-size 0.0005 both the estimation performance of ISS-LMS with step-size 0.005.

On the other hand, in the case of the same SNR regimes, when number of non-zero coefficient K becomes smaller, namely the channel becomes sparser, the estimation performance of proposed algorithms are improved as ISS-LMS for estimating sparse channel with step-size 0.0005, but faster than them.

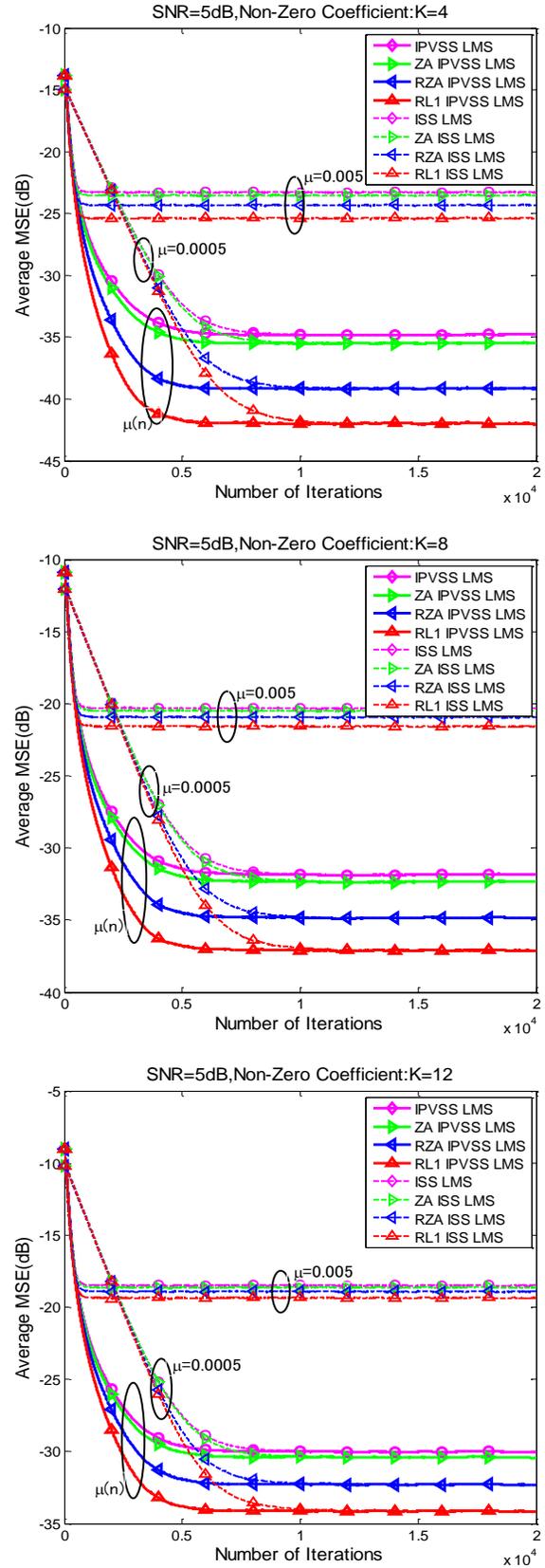

Fig. 4-6

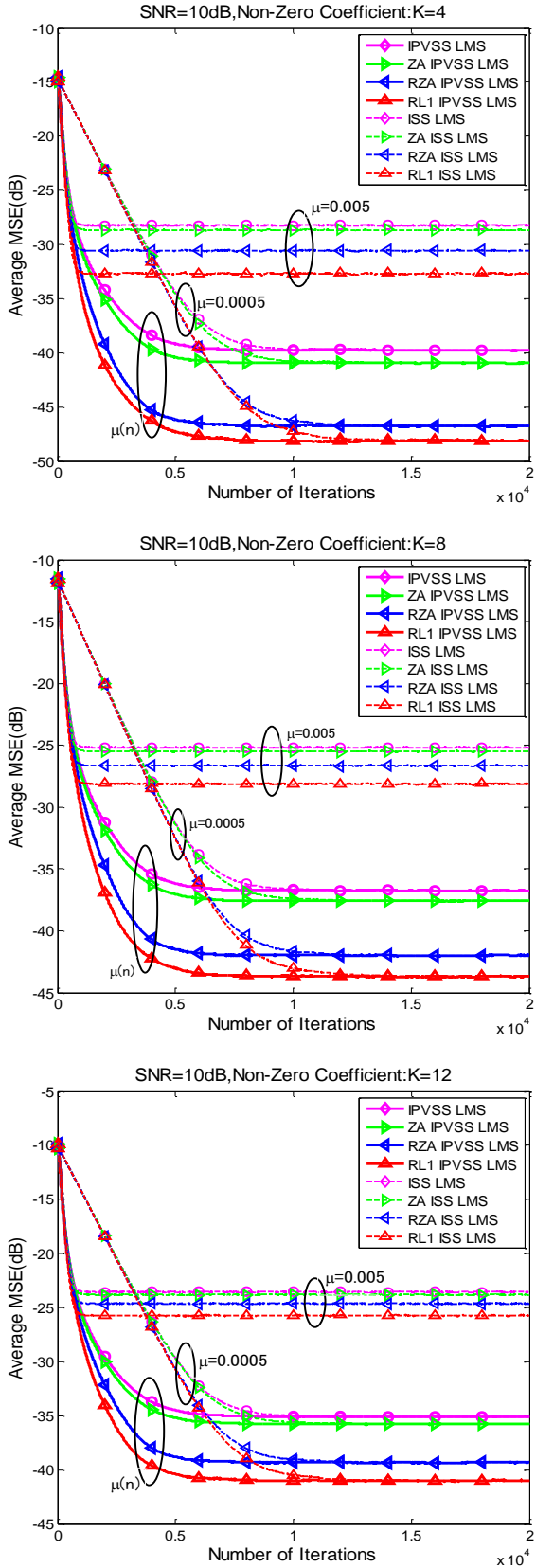

Fig. 7-9

As shown in Fig.10, we can control the estimation performance via hard threshold $\varphi$, compared with the algorithm with $\varphi = 0.0005$, another one with $\varphi = 0.0001$ can achieve a better estimation performance while the convergence speed is almost same.

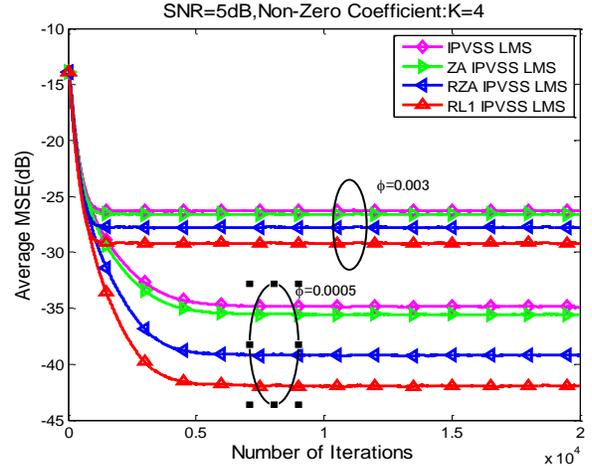

Fig. 10. MSE of different threshold $\varphi$ (SNR: 5db, K: 4)

## V. CONCLUSIONS

This paper has proposed three IPVSS-LMS algorithms to estimate sparse channels as well as balance between the estimation performance and the convergence speed. We first updated the equation of ZA-IPVSS-LMS, RZA-IPVSS-LMS and RL1-IPVSS-LMS from ZA-LMS, RZA-LMS and RL1-LMS. The performance enhancement is achieved via two aspects: VSS and channel sparsity. Compared to sparse ISS-LMS algorithms, our proposed VSS-LMS algorithms can further improve MSE performance while without scarifying convergence speed due to the fact that VSS is controlled by iteration as well as threshold. Simulation results were presented to validate the proposed channel estimation algorithms.